\documentclass{article}
\usepackage{arxiv}

\usepackage{amsmath,amsfonts,amssymb}
\usepackage{mathtools}
\usepackage[hidelinks]{hyperref}
\usepackage{algorithmic}
\usepackage{algorithm}
\usepackage{array}
\usepackage[caption=false,font=normalsize,labelfont=sf,textfont=sf]{subfig}
\usepackage{relsize}
\usepackage{textcomp}
\usepackage{stfloats}
\usepackage{url}
\usepackage{verbatim}
\usepackage{graphicx}
\usepackage{natbib}
\usepackage{doi}
\usepackage{xcolor}
\usepackage{paralist}
\usepackage{blindtext}
\usepackage{multicol}
\usepackage{multirow}
\usepackage[subnum]{cases}

\newtheorem{theorem}{Theorem}[section]

\newtheorem{remark}[theorem]{Remark}

\title{An Analytical Framework for Modeling and Synthesizing Malicious Attacks on ACC Vehicles}

\author{Shian Wang\\
	Electrical and Computer Engineering\\
	The University of Texas at El Paso\\
	\texttt{swang14@utep.edu}\\
}

\renewcommand{\shorttitle}{Modeling and Synthesizing Malicious Attacks on ACC Vehicles}

\begin{document}\sloppy
\maketitle

\begin{abstract}
    While emerging adaptive cruise control (ACC) technologies are making their way into more vehicles, they also expose a vulnerability to potential malicious cyberattacks. Previous research has typically focused on constant or stochastic attacks without explicitly addressing their malicious and covert characteristics. As a result, these attacks may inadvertently benefit the compromised vehicles, inconsistent with real-world scenarios. In contrast, we establish an analytical framework to model and synthesize a range of candidate attacks, offering a physical interpretation from the attacker's standpoint. Specifically, we introduce a mathematical framework that describes mixed traffic scenarios, comprising ACC vehicles and human-driven vehicles (HDVs), grounded in car-following dynamics. Within this framework, we synthesize and integrate a class of false data injection attacks into ACC sensor measurements, influencing traffic flow dynamics. As a first-of-its-kind study, this work provides an analytical characterization of attacks, emphasizing their malicious and stealthy attributes while explicitly accounting for vehicle driving behavior, thereby yielding a set of candidate attacks with physical interpretability. To demonstrate the modeling process, we perform a series of numerical simulations to holistically assess the effects of attacks on car-following dynamics, traffic efficiency, and vehicular fuel consumption. The primary findings indicate that strategically synthesized candidate attacks can cause significant disruptions to the traffic flow while altering the driving behavior of ACC vehicles in a subtle fashion to remain stealthy, which is supported by a series of analytical results.
\end{abstract}


\section{Introduction}\label{section1}

Although automated vehicles (AVs) are expected to revolutionize the transportation system by offering advantages such as lowered energy consumption~\citep{sun2022energy}, efficient parking space allocation~\citep{wang2021optimal}, and enhanced traffic flow~\citep{wang2022optimal}, the emergence of AV technologies also presents an opportunity for malicious actors to jeopardize vehicle safety and security~\citep{parkinson2017cyber}. Adaptive cruise control (ACC) vehicles, the first generation of AVs, are susceptible to cyberattacks in different forms, potentially leading to significant disruptions in regular traffic flow~\citep{petit2014potential}. Emerging connected autonomous vehicles (CAVs) are even more vulnerable to attacks as V2V communication channels are subject to additional adversarial actions~\citep{ju2022survey,han2023secure}.

Although cyberattacks may induce only subtle alterations in vehicle driving behavior~\citep{li2018influence}, these changes can lead to increased traffic congestion and energy usage~\citep{li2023exploring}. One typical type of attack that has garnered considerable research attention is the false data injection attack on sensor measurements~\citep{wang2020modeling,li2022detecting,ju2022survey,wang2023minmax,li2023exploring}, which can indirectly compel a vehicle to execute unintended maneuvers. In addition, intelligent vehicles are susceptible to various other types of malicious attacks, including attacks on vehicle control commands~\citep{li2022detecting,zhou2022robust,wang2023minmax,sun2023secure} and denial-of-service attacks~\citep{gao2022fixed,cai2023performance}. For a thorough exploration of potential cyberattacks on AVs, readers are directed to~\citep{petit2014potential}. The adverse effects of compromised ACC vehicles extend beyond their individual performance, affecting the overall dynamics of composite traffic comprising both ACC vehicles and human-driven vehicles (HDVs). For instance, even minor attacks on an individual vehicle could lead to decreased traffic efficiency and heightened energy usage, causing widespread disruption to the entire traffic flow~\citep{li2023exploring}. It is evident that potential attacks on ACC vehicles present a substantial risk to the safety, resilience and efficiency of the transportation system.

Although cyberattacks targeting intelligent vehicles have garnered growing interest within the transportation community in recent years, there remains a gap in research that systematically models and characterizes these attacks' malicious and covert attributes while explicitly taking into account fundamental car-following behavior. Previous research has predominantly relied on the assumption of constant or stochastic attacks against AVs or ACC vehicles~\citep{wang2021resilient,wang2020modeling,li2018influence,li2022detecting}. However, these assumptions may not adequately capture the malicious and covert characteristics of potential attacks. This is because such attacks poorly chosen by the adversary could even act in favor of an ACC vehicle, e.g., increasing ACC speed at a large spacing thereby improving throughput~\citep{li2022detecting}, contrary to their malicious nature. Moreover, modeling and characterizing potential attacks in this manner could lead to excessively aggressive driving behavior, potentially resulting in rear-end collisions~\citep{wang2020modeling}. Consequently, compromised vehicles are likely to be easily detected and identified using intrusion detection systems~\citep{khraisat2019survey} or anomaly detection techniques~\citep{li2022detecting}, not well aligned with the attacker's objective of maintaining the stealthiness of their attacks~\citep{gunter2021compromised}. 

As a result, questions on the realism of modeling and synthesizing cyberattacks on ACC vehicles in the way that has been presented in relevant studies naturally arise. To fill this fundamental research gap, we provide an analytical characterization of a range of malicious yet covert false data injection attacks targeting ACC sensor measurements based upon the understanding of vehicle driving behavior. Furthermore, we holistically study their effects on car-following dynamics, traffic efficiency, and vehicular energy consumption.

In summary, the main contributions of this work are:
\begin{itemize}
    \item We develop a comprehensive analytical framework designed to model and synthesize a broad class of false data injection attacks targeting ACC vehicles with a good level of realism. These attacks are executed using the true sensor measurements perceived by the attacker, without adhering to predetermined distributions. Notably, we consider not only attacks that can be launched additively but also multiplicatively, to ensure technical completeness.
    
    \item We carry out extensive mathematical analysis to characterize the malicious and covert attributes of possible attacks launched either additively or multiplicatively, based on the understanding of car-following driving behavior. This addresses the pressing issue of considering attacks as constants or Gaussian random variables incapable of describing their malicious and stealthy characteristics, as seen in many prior studies.
    
    \item As an extension of~\citep{wang2023novel}, we present more sophisticated analytical examples of candidate attacks and demonstrate their influence on traffic flow with a series of numerical simulations. 
    
    \item As an extension of~\citep{wang2023novel}, we derive detailed insights into the relationship between intelligently synthesized candidate attacks and car-following dynamics, traffic efficiency and vehicular energy consumption, which is anticipated to serve as a catalyst for future research endeavors aimed at developing effective strategies for detecting and mitigating attacks.
\end{itemize}

The subsequent sections of this article are structured as follows. Section~\ref{section2} introduces a mathematical framework for describing mixed traffic, with explicit discussion on car-following dynamics. Within the framework introduced, we mathematically model, synthesize, and analyze a wide range of covertly malicious attacks targeting ACC vehicles in Section~\ref{section3}, based on concrete car-following models. To demonstrate the modeling approach, we carry out comprehensive simulations in Section~\ref{section4} to assess the effects of candidate attacks on car-following dynamics, traffic efficiency, and vehicular energy usage. We conclude the article in Section~\ref{section5} discussing its limitations and possible future extensions. 


\section{Mathematical Model for Mixed Traffic}\label{section2}
\begin{figure}[t!]
	\centering
	\includegraphics[width=0.9\textwidth]{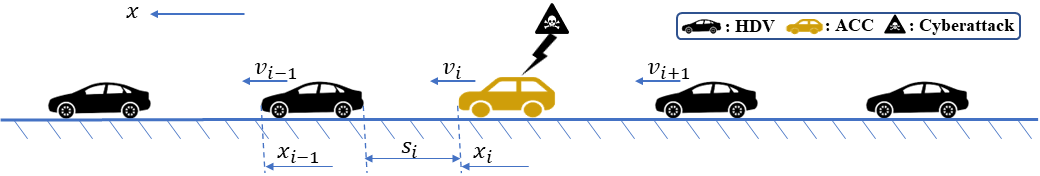}
	\caption{\textnormal{A graphic representation of mixed traffic consisting of ACC vehicles and HDVs, where ACC vehicles are vulnerable to cyberattacks while HDVs are not susceptible to such attacks.}}
	\label{cyberattack_mixed_traffic}
\end{figure}

We study the scenario of mixed traffic encompassing both HDVs and ACC vehicles (or AVs), a highly anticipated configuration in the foreseeable future~\citep{wang2022optimalTRC}. In line with previous research~\citep{wang2020modeling,li2022detecting,zhou2022robust,wang2023minmax,li2023exploring}, we limit our focus to the longitudinal vehicle dynamics, recognizing that lateral dynamics could also be explored. We consider a general string of $m$ vehicles represented by the ordered set $\mathcal{M} = \left\{ 1, 2, 3, \cdots, m \right\}$, where $m > 1$ and $m \in \mathbb{N}^{+}$. We signify the position and speed of vehicle $i \in \mathcal{M}$ at time $t$ as $x_{i}(t)$ and $v_{i}(t)$, respectively. The inter-vehicle spacing between vehicles $i$ and $i-1$ is defined as $s_{i}(t) = x_{i-1}(t) - x_{i}(t) - l_{i-1}$, where $l_{i-1}$ denotes the ($i-1$)th vehicle length. These notations are widely employed in the description of car-following dynamics~\citep{wilson2011car}, with a visual representation given in Figure~\ref{cyberattack_mixed_traffic}.

Based on the law of physics, the dynamics of any vehicle $i$ is written as~\citep{wilson2011car}
\begin{eqnarray}
    &~&\dot{x}_{i}(t) = v_{i}(t),    \label{eq2.1}   \\
    &~&\dot{v}_{i}(t) = f(s_{i}(t), \Delta v_{i}(t), v_{i}(t)), \label{eq2.2}
\end{eqnarray}
where the dot operator signifies differentiation with respect to time; the operator $f$ connects the acceleration of the $i$th vehicle, $\dot{v}_{i}$, with the variables $s_{i}$, $\Delta v_{i}$ and $v_{i}$; and the definition of the relative speed between vehicle $i$ and vehicle $i-1$ is
\begin{equation}
    \Delta v_{i}(t) = \dot{s}_{i}(t) = v_{i-1}(t) - v_{i}(t). \label{eq2.3}
\end{equation}
Equations~\eqref{eq2.1}--\eqref{eq2.2} embody a widely adopted functional expression for car-following dynamics. Although the variables, like $s_{i}$, $\Delta v_{i}$ and $v_{i}$, are time-dependent, we will, for the sake of brevity, omit the argument $t$ when applicable.

Let $\mathcal{H}$ and $\mathcal{A}$ represent the ordered sets of HDVs and ACC vehicles within the composite traffic, respectively. As human drivers often demonstrate distinct driving behaviors compared to ACC vehicles~\citep{ye2022car}, we differentiate $f$ in equation~\eqref{eq2.2} as $f_{\text{HDV}}$ and $f_{\text{ACC}}$ for HDVs and ACC vehicles, respectively. Hence, it follows that
\begin{equation}
    \dot{x}_{i} = v_{i}, ~\forall~ i \in \mathcal{M} = \mathcal{H} \cup \mathcal{A}   \label{eq2.4}
\end{equation}
\begin{numcases}{\dot{v}_{i} =}
    f_{\text{HDV}}(s_{i}, \Delta v_{i}, v_{i}), ~\forall~ i \in \mathcal{H}    \label{eq2.5a}  \\  
    f_{\text{ACC}}(s_{i}, \Delta v_{i}, v_{i}), ~\forall~ i \in \mathcal{A}    \label{eq2.5b}
\end{numcases}
It is important to emphasize that $f_{\text{HDV}}$ and $f_{\text{ACC}}$ do not have to be same. Rather, they rely on the particular car-following principles adhered to by HDVs and ACC vehicles. Here we do not incorporate reaction-time delays in equations~\eqref{eq2.5a}--\eqref{eq2.5b}, consistent with previous work modeling traffic flow based on car-following dynamics~\citep{talebpour2016influence,wang2020modeling,li2022detecting,wilson2011car,wang2022optimal,wang2023general}.

Equations~\eqref{eq2.5a}--\eqref{eq2.5b} describe vehicle driving behavior in terms of car-following dynamics. To ensure the consistency between a car-following model, e.g., equation~\eqref{eq2.2}, and its corresponding practical driving behavior, the subsequent rational driving constraints~(RDC)~\citep{wilson2011car} are expected to be met
\begin{eqnarray}
    \beta_{1} \coloneqq \frac{\partial \dot{v}}{\partial s} > 0, ~~ \beta_{2} \coloneqq \frac{\partial \dot{v}}{\partial \Delta v} > 0, ~~ \beta_{3} \coloneqq \frac{\partial \dot{v}}{\partial v} < 0. \label{eq2.6}
\end{eqnarray}
The expressions in~\eqref{eq2.6} can be interpreted as follows~\citep{wilson2011car}: $\beta_{1} > 0$ suggests that a larger spacing should lead to more acceleration; $\beta_{2} > 0$ indicates that a greater relative speed should result in increased acceleration; $\beta_{3} < 0$ implies that a vehicle tends to accelerate less as its speed increases. The RDC provides a straightforward criterion for the presence of rational car-following models. Interested readers may refer to~\citep{wilson2011car} for a detailed interpretation on the expressions shown in~\eqref{eq2.6}.

\section{Modeling and Synthesizing Cyberattacks on Adaptive Cruise Control Vehicles}\label{section3}

Although cyberattacks can manifest in various forms against ACC vehicles, our exclusive focus is on false data injection attacks targeting ACC sensor measurements, a type of attack garnering substantial research attention in recent years~\citep{wang2020modeling,li2022detecting,wang2023minmax,li2023exploring}. Specifically, we develop an analytical framework for modeling and synthesizing these attacks within the context of car-following dynamics, with ACC vehicles being potentially vulnerable targets facing adversaries, illustrated in Figure~\ref{cyberattack_mixed_traffic}. While ACC vehicles could be vulnerable to cyberattacks, it is considered that HDVs are not susceptible to such attacks as seen in prior studies. Previous work has mostly considered the assumption of constant attacks or stochastic attacks with a known statistical distribution, targeting AVs or ACC vehicles~\citep{wang2021resilient,wang2020modeling,li2018influence,li2022detecting}. As discussed before, this may not correctly capture the malicious covert traits of attackers. We will integrate attacks on ACC sensor measurements into the traffic dynamics presented before, with a demonstration leveraging concrete car-following models. Additionally, we provide an analytical characterization of the malicious and covert attributes of these candidate attacks, grounded in a physically interpretable framework rooted in the understanding of car-following behavior.

\subsection{A general framework for mixed traffic under cyberattacks}\label{section3_1}

The desired acceleration of an ACC vehicle, given by equation~\eqref{eq2.5b}, depends on its sensor measurements, including $s_i$ and $\Delta v_i$. These measurements could be compromised by cyberattacks corrupting ACC hardware sensors or the software algorithms used for data acquisition, in the form of false data injection attacks~\citep{wang2020modeling,li2022detecting,wang2022planning,wang2023minmax,li2023exploring}. For any vehicle $i \in \mathcal{A}$, let $\omega_{1,i}$ and $\omega_{2,i}$ represent the false data injection attacks on spacing and relative speed, respectively. Hence, the resulting acceleration of any ACC vehicle $i$ under attack, $i \in \mathcal{A}$, is
\begin{eqnarray}
\dot{v}_{i} = f_{\text{ACC}}(s_{i}+\omega_{1,i}, \Delta v_{i}+\omega_{2,i}, v_{i}) \coloneqq F(s_{i}, \Delta v_{i}, v_{i}, \omega_{i}),   \label{eq3.1}
\end{eqnarray}
with $\omega_{i} = \left[ \omega_{1,i}, \omega_{2,i} \right]$ signifying the attack vector on ACC sensor measurements. While attacks are incorporated into ACC dynamics in an additive manner, commonly seen in many relevant studies~\citep{wang2020modeling,li2022detecting,wang2022planning,li2023exploring}, we will also briefly discuss about modeling candidate attacks in a multiplicative fashion. 

As a result of false data injection attacks, the mixed traffic dynamics describing vehicle acceleration is given by
\begin{numcases}{\dot{v}_{i} =}
    f_{\text{HDV}}(s_{i}, \Delta v_{i}, v_{i}), ~\text{if}~ i \in \mathcal{H},  \label{eq3.2a}  \\
    f_{\text{ACC}}(s_{i}, \Delta v_{i}, v_{i}), ~\text{if}~ i \in \mathcal{A} ~\text{not attacked},    \label{eq3.2b}  \\ 
    F(s_{i}, \Delta v_{i}, v_{i}, \omega_{i}), ~\text{if}~ i \in \mathcal{A} ~\text{attacked}.   \label{eq3.2c}
\end{numcases}

Our primary objective is to mathematically model and characterize candidate attacks so that their adversarial and stealthy nature can be captured in a physically interpretable manner. This will allow for a higher degree of realism compared to many prior studies assuming constant attacks or random attacks like Gaussian noise without explicitly characterizing the intention of malicious actors. Being able to synthesize attacks with greater realism provides a more profound understanding of their real-world implications on traffic dynamics, thereby facilitating the development of effective and robust strategies for attack detection and mitigation.

It is reasonable to assume that false data injection attacks are initiated using the true ACC sensor measurements as perceived by the attacker. Namely, $\omega_{1,i}$ is considered a function of the actual spacing $s_i$, while $\omega_{2,i}$ is represented by a function of the relative speed $\Delta v_i$. This is mathematically written as
\begin{eqnarray}
    \omega_{1,i} = g_1(s_i), ~ \omega_{2,i} = g_2(\Delta v_i), \label{eq3.3}
\end{eqnarray}
where $g_1$ and $g_2$ are differentiable with respect to the sensor measurements $s_i$ and $\Delta v_i$, respectively. Clearly, this covers a range of false data injection attacks that could potentially occur to ACC sensor measurements. It is worth noting that the functions, $g_1$ and $g_2$, likely take nonzero values only for a specific period of time. That is, an attacker does not have to consistently or continuously launch attacks; they might do so sporadically or intermittently.

\subsection{Illustration of mixed traffic under cyberattacks}\label{section3_2}

Building upon the framework introduced before, we proceed to provide a practical illustration of mixed traffic flow dynamics using specific car-following models, facilitating mathematical analysis and numerical investigations. The intelligent driver model
(IDM) is utilized for HDVs due to its proven accuracy in depicting the behavior of human drivers with commendable performance~\citep{treiber2000congested,talebpour2016influence}. The IDM is a multi-regime model capable of offering a high level of realism in representing various congestion levels~\citep{sarker2019review}. Furthermore, recent research has demonstrated the IDM's ability to closely emulate human driving behavior with exceptional accuracy, outperforming other car-following models in real-world driving data assessments~\citep{kesting2008calibrating,pourabdollah2017calibration,he2023calibrating}.

According to the IDM, for $i \in \mathcal{H}$ equation~\eqref{eq3.2a} is given by
\begin{eqnarray}
    &~&\hskip-20pt f_{\text{HDV}} = a \left[ 1 - \left(\frac{v_{i}}{v_{0}}\right)^{4}  - \left(\frac{s^{\ast}(v_{i},\Delta v_{i})}{s_{i}}\right)^{2} \right],   \label{eq3.4}
\end{eqnarray}
with
\begin{equation}
    s^{\ast}(v_{i},\Delta v_{i}) = s_{0} + \max\left\{0, v_{i}T - \frac{v_{i}\Delta v_{i}}{2\sqrt{ab}}\right\},    \label{eq3.5}
\end{equation}
where $a$ is the maximum acceleration, $b$ represents the comfortable braking deceleration, $v_{0}$ signifies the desired speed, $s_{0}$ denotes the minimum spacing, and $T$ indicates the desired time headway.

As noted in~\citep{ye2022car}, ACC vehicles often exhibit distinct driving behaviors compared to human drivers. While ACC vehicles from different manufacturers may feature varying acceleration functions described in equation~\eqref{eq2.5b}, we adopt the optimal velocity with relative velocity (OVRV) model~\citep{milanes2013cooperative} to describe ACC car-following behavior for multiple reasons. The OVRV model is extensively used in the context of ACC systems~\citep{milanes2013cooperative}. It adheres to a constant time-gap policy, which aligns with the implementation of intelligent vehicles~\citep{ioannou1993autonomous}. Furthermore, the OVRV model has demonstrated its capability to accurately represent both simulated and actual vehicle trajectories involving ACC vehicles~\citep{milanes2013cooperative}.

According to the OVRV model, equation~\eqref{eq3.2b} is given by
\begin{eqnarray}
    f_{\text{ACC}} = k_{1}\left(s_{i} - \eta - \tau v_{i}\right) + k_{2}\Delta v_{i}, ~ i \in \mathcal{A}, ~ \text{not attacked},       \label{eq3.6}
\end{eqnarray}
where $\eta$ denotes the jam distance, i.e., inter-vehicle spacing at rest, $\tau$ represents the desired time gap, and $k_{1}$ and $k_{2}$ are positive parameters on the time gap and relative speed, respectively. 

It follows from the general framework presented above that the acceleration dynamics, $F$, for an ACC vehicle under attack on sensor measurements, as defined by equation~\eqref{eq3.2c}, becomes
\begin{eqnarray}
    F = k_{1}\left(s_{i} + \omega_{1,i} - \eta - \tau v_{i}\right) + k_{2}\left(\Delta v_{i} + \omega_{2,i}\right), ~ i \in \mathcal{A}, ~ \text{attacked}.      \label{eq3.7}
\end{eqnarray}

\subsection{Mathematical characterization of malicious attacks}\label{section3_3}

Here we characterize the mathematical properties of $\omega_{1,i}$ and $\omega_{2,i}$ for attacks to remain stealthy in relation to car-following dynamics. Specifically, we introduce the following physically interpretable conditions based upon the understanding of vehicle driving behavior and attacker intentions. 
\begin{itemize}
    \item[(i)] The rational driving behavior, defined by the RDC in~\eqref{eq2.6}, shall be preserved in attacked vehicles, as irrational driving patterns could be promptly detected and recognized.
    \item[(ii)] Collision shall be avoided for attacked vehicles, as viewed from the attacker's standpoint, to prevent the compromised vehicles from being easily identified~\citep{gunter2021compromised}.    
\end{itemize}
Machine learning methods can be employed for anomaly detection in intelligent vehicles~\citep{khanapuri2021learning,he2023wkn}. For instance, the recent study~\citep{li2022detecting} has shown that commonly used random attacks, represented by Gaussian noise, on ACC vehicles can be effectively detected via machine learning based generative adversarial networks. It is also reasonable to assume that erratic, uncalculated and aggressive attacks can be efficiently detected by intrusion detection systems~\citep{khraisat2019survey}. Hence, for a good degree of realism it is necessary to synthesize candidate attacks taking into account the driving behavior of vehicles, which has been largely ignored in previous research. Essentially, the conditions (i) and (ii) presented above indicate that attackers tend to alter the driving behavior of compromised ACC vehicles in a subtle manner for remaining stealthy. Nevertheless, even minor alterations in vehicle driving behavior have the potential to trigger extensive disruptions within the transportation network, leading to significant traffic congestion and heightened energy consumption and emissions from vehicles~\citep{li2023exploring}.

We first examine the mathematical implications of condition~(i) on potential attacks shown in equation~\eqref{eq3.3}. To ensure the RDC, it follows from the attacked ACC dynamics of equation~\eqref{eq3.2c} that
\begin{eqnarray}
    &~& \tilde{\beta}_{1} \coloneqq {\partial F}/{\partial s} > 0,  \label{eq3.8a}  \\
    &~& \tilde{\beta}_{2} \coloneqq {\partial F}/{\partial \Delta v} > 0,    \label{eq3.8b}  \\ 
    &~& \tilde{\beta}_{3} \coloneqq {\partial F}/{\partial v} < 0,   \label{eq3.8c}
\end{eqnarray}
where the subscript $i$ is dropped for brevity. Using the chain rule, the above expression~\eqref{eq3.8a} is explicitly given by
\begin{eqnarray}
    \tilde{\beta}_{1} = \frac{d}{ds}f_{\text{ACC}}\bigg|_{s=s+g_1} \cdot \frac{d}{ds}\left(s+g_1\right) = k_{1}\left(1+g_{1}'\right) > 0,       \label{eq3.9}
\end{eqnarray}
where $f_{\text{ACC}}$ corresponds to the ACC dynamics in normal operation (i.e., in the absence of attack) as introduced in equation~\eqref{eq3.6}. Similarly, it follows from~\eqref{eq3.8b} and~\eqref{eq3.8c} that 
\begin{eqnarray}
    \tilde{\beta}_{2} = k_{2}\left(1+g_{2}'\right) > 0, ~~ \tilde{\beta}_{3} = \beta_{3} = -\tau k_1 < 0.  \label{eq3.10}
\end{eqnarray}
Since $k_{1}, k_{2} > 0$, it is clear that
\begin{eqnarray}
    1 + g_1' > 0, ~~ 1 + g_2' > 0,   \label{eq3.11}
\end{eqnarray}
yielding the following conditions
\begin{eqnarray}
    g_1' > -1, ~~ g_2' > -1.    \label{eq3.12}
\end{eqnarray}

Now we consider the condition~(ii) to derive the corresponding mathematical properties of stealthy attacks. Based upon the comprehension of car-following dynamics, a rear-end collision could occur when an attack is launched in such a way that the spacing or relative speed perceived by the attacked vehicle is larger than its actual value. For example, it follows from equation~\eqref{eq3.1} and the first expression of~\eqref{eq2.6} that an attacked vehicle tends to accelerate more aggressively if the perceived spacing, $s_{i}+\omega_{1,i}$, is greater than the actual spacing, $s_{i}$. In other words, the attacked vehicle is spoofed to speed up while there is actually not sufficient space for it to do so safely. A similar logic easily follows for attack on relative speed. Hence, to reduce the likelihood of collision due to malicious attacks, one may impose the following conditions
\begin{eqnarray}
   \frac{\partial F}{\partial s} \leq \frac{\partial f_{\text{ACC}}}{\partial s},   \label{eq3.13} 
\end{eqnarray}
\begin{eqnarray}
   \frac{\partial F}{\partial \Delta v} \leq \frac{\partial f_{\text{ACC}}}{\partial \Delta v},   \label{eq3.14}
\end{eqnarray}
for an attacked ACC vehicle, omitting the subscript $i$ for conciseness. Expression~\eqref{eq3.13} signifies that when the inter-vehicle spacing $s$ increases, an attacked vehicle exhibits less proactive driving behavior, whereas it adopts a more conservative approach when the spacing decreases, in contrast to normal ACC operation. Consequently, this is likely to degrade the efficiency of traffic flow, characterizing the malicious nature of adversaries. The inequality~\eqref{eq3.14} indicates that the attacked vehicle displays a heightened reactive response to changes in relative speed $\Delta v$. It adopts a more conservative approach when its speed aligns with that of the preceding vehicle, as opposed to the typical behavior in ACC dynamics. Clearly, such characterization of candidate attacks could result in relatively reactive driving behavior of a compromised vehicle, thereby reducing the risk of collision for attacks to remain stealthy. Unfortunately, it will not suffice for achieving this by simply employing a stochastic process, e.g., Gaussian noise~\citep{li2022detecting}, to describe potential cyberattacks. This is because random attacks may easily result in a rear-end collision. Moreover, they could even act in favor of compromised ACC vehicles in many scenarios, which is inconsistent with their adversarial intention. Using the ``$\leq$'' sign in the expressions of~\eqref{eq3.13} and~\eqref{eq3.14} aligns with the notion that, from the attacker's standpoint, collision shall be prevented, or reduced at least, in order to maintain stealthiness.

It follows from the expressions~\eqref{eq3.13} and~\eqref{eq3.14} that 
\begin{eqnarray}
    &~& \tilde{\beta}_{1} = k_{1}\left(1+g_{1}'\right) \leq k_{1},   \label{eq3.15}   \\
    &~& \tilde{\beta}_{2} = k_{2}\left(1+g_{2}'\right) \leq k_{2},   \label{eq3.16}
\end{eqnarray}
leading to the following conditions
\begin{eqnarray}
    g_1' \leq 0, ~~ g_2' \leq 0.    \label{eq3.17}
\end{eqnarray}

It follows from the conditions of~\eqref{eq3.12} and~\eqref{eq3.17} that
\begin{eqnarray}
    -1 < g_1' \leq 0, ~ -1 < g_2' \leq 0.    \label{eq3.18}
\end{eqnarray}

It is worth noting that the above conditions shown in~\eqref{eq3.18} can be slightly relaxed to allow for $g_1' = g_2' = -1$. In other words, it is possible to have $\tilde{\beta}_{1} = \tilde{\beta}_{2} = 0$, which essentially indicates that the driving behavior of an ego ACC vehicle is independent of the vehicle ahead at large spacings. Interested reader is referred to~\citep{wilson2011car} for a detailed interpretation. Considering $\mathcal{C}$ as the set of additive malicious attacks, it follows that
\begin{eqnarray}
    \mathcal{C} \coloneqq \left\{g_1, g_2: -1 \leq g_1' \leq 0, ~ -1 \leq g_2' \leq 0\right\}. \label{eq3.19}
\end{eqnarray}

\begin{remark}\label{remark3.1}
While we do not intend to exhaustively synthesize all possible attacks, the set $\mathcal{C}$ encompasses a wide range of malicious yet covert candidate attacks targeting ACC vehicles. It is worth noting that these attacks are launched to ACC sensor measurements in an additive manner as seen from equation~\eqref{eq3.1}. While attacks could also originate from outside the set $\mathcal{C}$, they might induce more erratic driving behavior in attacked vehicles, potentially leading to rear-end collisions, which could increase the likelihood of detection and identification.
\end{remark}

While attacks are commonly modeled in an additive fashion as presented so far, it is not exactly clear how real-world attacks could corrupt the sensor measurements of ACC vehicles from a mathematical and analytical standpoint. Thus, for technical completeness we also provide a brief discussion on incorporating multiplicative attacks into the dynamics of ACC vehicles. To this end, the effective acceleration dynamics of any ACC vehicle $i$ under attack, $i \in \mathcal{A}$, is given by 
\begin{eqnarray}
\dot{v}_{i} = f_{\text{ACC}}(s_{i} \cdot g_1(s_{i}), \Delta v_{i} \cdot g_2(\Delta v_{i}), v_{i}) \coloneqq F(s_{i}, \Delta v_{i}, v_{i}, \omega_{i}),   \label{eq3.20}
\end{eqnarray}
where $\omega_{i} = \left[ \omega_{1,i}, \omega_{2,i} \right] = \left[ g_1(s_{i}), g_2(\Delta v_{i}) \right]$ is the multiplicative attack vector on ACC sensor measurements. 

It follows from the expressions~\eqref{eq3.8a}--\eqref{eq3.8b} ensuring RDC that 
\begin{eqnarray}
    \tilde{\beta}_{1} = \frac{d}{ds}f_{\text{ACC}}\bigg|_{s=s \cdot g_1} \cdot \frac{d}{ds}\left(s \cdot g_1\right) = k_{1}\left(g_{1} + s \cdot g_{1}'\right) > 0       \label{eq3.21}
\end{eqnarray}
and
\begin{eqnarray}
    \tilde{\beta}_{2} = k_{2}\left(g_{2} + \Delta v \cdot g_{2}'\right) > 0.       \label{eq3.22}
\end{eqnarray}

Similarly, it follows from~\eqref{eq3.13} and~\eqref{eq3.14} that 
\begin{eqnarray}
    g_{1} + s \cdot g_{1}' \leq 1, ~~ g_{2} + \Delta v \cdot g_{2}' \leq 1.   \label{eq3.23}
\end{eqnarray}
Hence, the set of potential multiplicative malicious attacks is given by
\begin{eqnarray}
    \mathcal{D} \coloneqq \left\{g_1, g_2: 0 < g_{1} + s \cdot g_{1}' \leq 1, 0 < g_{2} + \Delta v \cdot g_{2}' \leq 1 \right\}.    \label{eq3.24}
\end{eqnarray}

\begin{remark}\label{remark3.2}
Multiplicative attacks present an additional way for an attacker to compromise the sensor measurements of an ACC vehicle. The formulation remains consistent with the assumption that attacks are initiated using the actual ACC sensor measurements as perceived by the attacker. 
\end{remark}

\subsection{Illustrative examples of candidate attacks}\label{section3_4}

Since the present study focuses primarily on additive attacks commonly seen in the literature, we first show a concrete example of candidate attacks from the set $\mathcal{C}$, followed by a discussion about attacks with slightly relaxed assumptions on differentiability and continuity with respect to ACC sensor measurement. Then, we present some brief illustrative results on attacks that could be launched in a multiplicative manner. As an extension of~\citep{wang2023novel} which was significantly limited to linear functions of $g_1$ and $g_2$, in the present study we allow for any feasible differentiable function (linear or nonlinear), with concrete examples presented below.

The set $\mathcal{C}$ covers a range of candidate attacks, with a simple example, say \textit{Example~1}, given by linear functions of ACC sensor measurement as follows
\begin{eqnarray}
    &~& \omega_{1} = g_1(s) = r_1s,   \label{eq3.25}   \\
    &~& \omega_{2} = g_2(\Delta v) = r_2\Delta v,  \label{eq3.26}
\end{eqnarray}
where $r_1, r_2 \in \left[-1, 0\right]$. However, these candidate attacks could also exhibit a nonlinear relationship with sensor measurement. One possible example, say \textit{Example~2}, of such attacks is
\begin{eqnarray}
    &~& \omega_{1} = g_1(s) = 0.4\sin(s) - 0.5s,   \label{eq3.27}   \\
    &~& \omega_{2} = g_2(\Delta v) = 0.4\sin(\Delta v) - 0.5\Delta v.  \label{eq3.28}
\end{eqnarray}
It is easy to verify that $g_{1}', g_{2}' \in \left[-0.9, -0.1\right]$, hence, $g_{1}, g_{2} \in \mathcal{C}$. It is worth mentioning that the set $\mathcal{C}$ is derived with the understanding that $g_1$ and $g_2$ are differentiable with respect to their arguments, $s$ and $\Delta v$, respectively. While almost everywhere continuous differentiable functions of $g_1$ and $g_2$ may not satisfy the expressions of~\eqref{eq3.8a}--\eqref{eq3.8b} due to nonexistence of derivatives at certain points, they may still compromise ACC vehicles in a stealthy manner as long as such attacks do not alter vehicle driving behavior in an irrational fashion.

Now we briefly present an example, say \textit{Example~3}, of candidate attacks launched in a multiplicative manner (shown in equation~\eqref{eq3.20}) characterized by the set $\mathcal{D}$. Taking the attack on inter-vehicle spacing, i.e., $g_1$, for illustration, it follows from the condition of~\eqref{eq3.21} that
\begin{eqnarray}
    g_{1} + s \cdot g_{1}' > 0.       \label{eq3.29}
\end{eqnarray}
Since we are deriving a candidate attack $g_1 \in \mathcal{D}$, one can assume that $g_1 > 0$. Consequently, the above inequality~\eqref{eq3.29} is equivalent to
\begin{eqnarray}
    (1/g_{1})dg_{1} > -(1/s)ds.        \label{eq3.30}
\end{eqnarray}
Integrating both sides of~\eqref{eq3.30} and carrying out basic calculation leads to
\begin{eqnarray}
    \ln(s \cdot g_1) > 0,        \label{eq3.31}
\end{eqnarray}
which implies
\begin{eqnarray}
    g_1 > 1/s.        \label{eq3.32}
\end{eqnarray}
Thus, one can define $g_1 = 1/s + z$ with $z > 0$ being a constant. It follows from the first inequality of~\eqref{eq3.23} conditioning on $g_1$ that 
\begin{eqnarray}
    g_1 + s \cdot \frac{d}{ds}\left(1/s + z\right) \leq 1,        \label{eq3.33}
\end{eqnarray}
which results in
\begin{eqnarray}
    z \leq 1.        \label{eq3.34}
\end{eqnarray}
Hence, a multiplicative candidate attack of $g_1$ can be introduced as follows
\begin{eqnarray}
    g_1 = 1/s + z, ~ z \in \left(0, 1\right].       \label{eq3.35}
\end{eqnarray}
Similarly, one can derive candidate attacks of $g_2$ that could be launched in a multiplicative manner, following the conditions shown in~\eqref{eq3.24}.

\begin{remark}\label{remark3.3}
It is noted that, when being launched in a multiplicative manner, attacks can involve not only the sensor measurement but also additional constant terms like $z$ seen in equation~\eqref{eq3.35}. 
\end{remark}

\begin{remark}\label{remark3.4}
From the attacker perspective, one may tend to maximize their impact on traffic flow by launching select attacks from the set derived. This can be properly formulated as an optimization problem with the conditions on attacks, e.g., the set $\mathcal{C}$ in equation~\eqref{eq3.19}, serving as constraints. In fact, for the special case with $g_1$ and $g_2$ being linear functions it is observed that attacks drawn from the set $\mathcal{C}$ could have maximal disruption to vehicle speed when they lie on the boundary of $\mathcal{C}$, i.e., $g_{1}' = g_{2}' = -1$. This is because the compromised vehicle is spoofed to the greatest extent as understood from the car-following dynamics shown in equation~\eqref{eq3.1}. We will further comment on this in the subsequent section. However, such an optimization problem could be challenging to solve in general due to the fact that $g_1$ and $g_2$ are functional of time (not simply a function of time $t$). Since this is beyond the scope of the present work, we leave it for future study. 
\end{remark}

\section{Numerical Results}\label{section4}

Here we perform numerical simulations using Matlab to demonstrate the analytical findings from the previous sections. As introduced before, we utilize the IDM and OVRV model to represent HDVs and ACC vehicles, respectively. The parameter values for these models are summarized in Tables~\ref{Table_parameters_IDM} and~\ref{Table_parameters_OVRV}, with the IDM parameters being standard for freeway traffic~\citep{treiber2013traffic}. The OVRV model parameters are sourced from~\citep{gunter2019modeling}, based upon realistic data collected from commercially available ACC vehicles~\citep{gunter2020commercially}.

\begin{figure}[t!]
	\centering
	\includegraphics[width=0.9\textwidth]{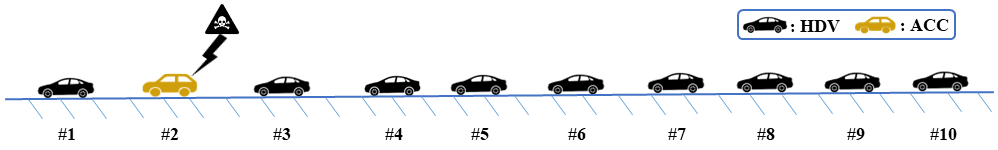}
	\caption{\textnormal{A graphic illustration depicting a 10-vehicle string in mixed traffic, where the second vehicle is an ACC vehicle susceptible to potential attacks.}}
	\label{Simulation_setting}
\end{figure}

For numerical illustration, we focus primarily on candidate attacks launched in an additive manner shown in equation~\eqref{eq3.1}, i.e., from the set $\mathcal{C}$, as commonly considered in the literature. It is worth mentioning that no inter-vehicle communication is required for an ACC system, since the inputs for an ACC vehicle, i.e., relative speed and spacing to its preceding vehicle, can be readily obtained through onboard sensors. Hence, we consider an ACC vehicle under attack following a lead vehicle. In addition, we assess how the attacked vehicle affects the bulk traffic, specifically its impact on the following vehicles. Specifically, we analyze a generic scenario featuring a string of 10 vehicles as shown in Figure~\ref{Simulation_setting}, where the second vehicle is an ACC vehicle susceptible to potential attacks. It is not necessary to consider variation of ACC market penetration rates since no cooperation is required among ACC vehicles and attacks are launched to individual vehicles. However, it is reasonable to speculate that a larger presence of compromised vehicles may result in a greater collective impact on traffic flow. Similar to the settings of many relevant studies~\citep{wang2020modeling,li2022detecting,zhou2022robust,wang2023minmax}, we exclusively focus on longitudinal vehicle dynamics, although lateral dynamics could also be explored. In what follows, we examine the impact of candidate attacks, taken from both the set $\mathcal{C}$ and its complement, on car-following dynamics. Then, we evaluate their effects on traffic performance pertaining to average speed variation and vehicular energy consumption. 

\begin{table}[t!]
\setlength{\tabcolsep}{5pt}
\caption{\textnormal{Parameter Values Used for the IDM~\citep{treiber2013traffic}}}
\label{Table_parameters_IDM}
\begin{center}
 \begin{tabular}{c c c c}
 \hline
 \textbf{Parameter} ~&~ \textbf{Description} ~&~ \textbf{Value}  \\  [0.5ex]
 \hline
 $v_{0}$ ~&~ desired speed (\text{m/s}) ~&~ 30.0  \\ [0.3ex]
 \hline
 $T$ ~&~ time gap (\text{s}) ~&~ 1.5  \\  [0.3ex]
 \hline
 $s_{0}$ ~&~ minimum spacing (\text{m}) ~&~ 2.0  \\[0.3ex]
 \hline
 $a$ ~&~ maximum acceleration (\text{m/s\textsuperscript{2}}) ~&~ 1.4   \\  [0.3ex]
 \hline
 $b$ ~&~ comfortable deceleration (\text{m/s\textsuperscript{2}}) ~&~ 2.0   \\  [0.3ex]
 \hline
 $l_{i}$ ~&~ vehicle length (\text{m}) ~&~ 5.0   \\ [0.3ex]
 \hline
\end{tabular}
\end{center}
\end{table}

\begin{table}[t!]
\setlength{\tabcolsep}{5pt}
\caption{\textnormal{Parameter Values Used for the OVRV Model~\citep{gunter2019modeling}}}
\label{Table_parameters_OVRV}
\begin{center}
 \begin{tabular}{c c c c}
 \hline
 \textbf{Parameter} ~&~ \textbf{Description} ~&~ \textbf{Value}  \\  [0.5ex]
 \hline
 $k_{1}$ ~&~ gain on the time gap (\text{s\textsuperscript{-2}}) ~&~ 0.02   \\  [0.3ex]
 \hline
 $k_{2}$ ~&~ gain on the relative speed (\text{s\textsuperscript{-1}}) ~&~ 0.13  \\  [0.3ex]
 \hline
 $\eta$ ~&~ jam distance (\text{m}) ~&~ 21.51  \\  [0.3ex]
 \hline
 $\tau$ ~&~ desired time gap (\text{s}) ~&~ 1.71  \\ [0.3ex]
 \hline
 $l_{i}$ ~&~ vehicle length (\text{m}) ~&~ 5.0   \\ [0.3ex] 
 \hline
\end{tabular}
\end{center}
\end{table}

\subsection{Impact of candidate attacks on car-following dynamics}\label{section4_1}

\begin{figure}[t!]
  \centering
  \subfloat[Case~1: $g_1 = 0.1\sin(s) - 0.1s$, \\ $g_2 = 0.1\sin(\Delta v) - 0.1\Delta v$]{\includegraphics[width=0.33\textwidth]{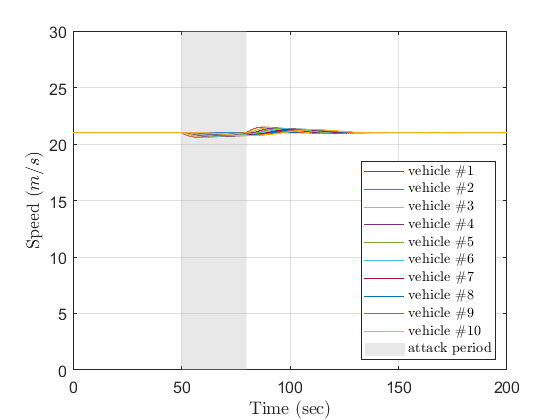}\label{speed_case1}}
  \hfil%
  \subfloat[Case~2: $g_1 = 0.4\sin(s) - 0.5s$, \\ $g_2 = 0.4\sin(\Delta v) - 0.5\Delta v$]{\includegraphics[width=0.33\textwidth]{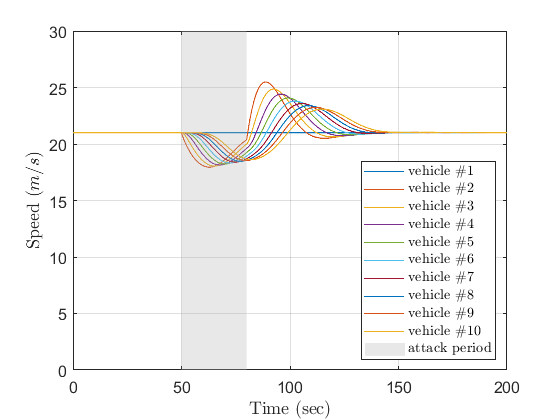}\label{speed_case2}}
  \hfil%
  \subfloat[Case~3: $g_1 = - 0.9s$, $g_2 = - 0.9\Delta v$]{\includegraphics[width=0.33\textwidth]{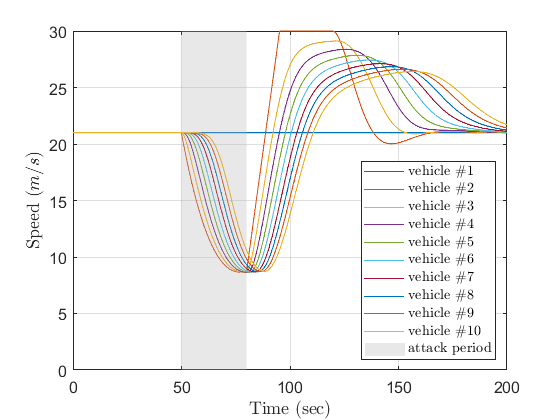}\label{speed_case3}}
  \vfil%
  \subfloat[Case~4: $g_1 = - 2s$, $g_2 = - 2\Delta v$]{\includegraphics[width=0.33\textwidth]{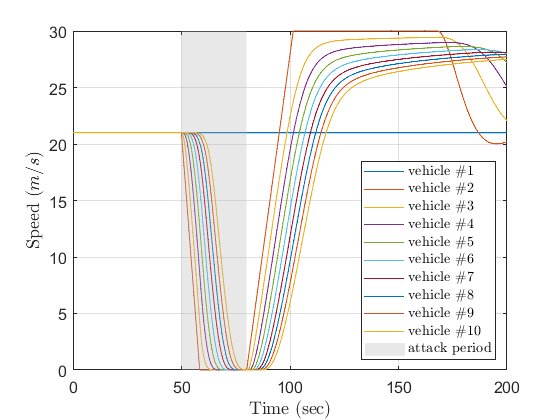}\label{speed_case4}}
  \hfil%
  \subfloat[Case~5: $g_1 = 10s$, $g_2 = - 5\Delta v$]{\includegraphics[width=0.33\textwidth]{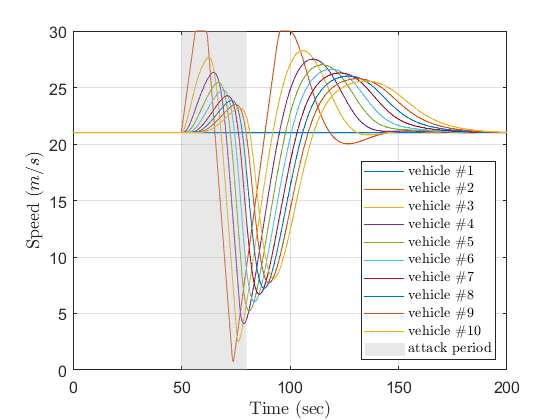}\label{speed_case5}}
  \hfil%
  \subfloat[Case~6: $g_1 = \sin(s) + 20s$, \\ $g_2 = \sin(\Delta v) - 20\Delta v$]{\includegraphics[width=0.33\textwidth]{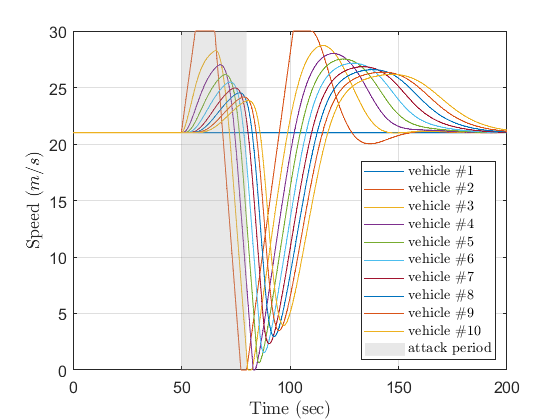}\label{speed_case6}}  
  \caption{\textnormal{Speed profile of all vehicles for Case~1 -- Case~6, with attacks occurring during the period of $I = \left[50, 80\right]$~sec. The exact form of candidate attacks for each case is given in its corresponding figure caption, e.g., $g_1 = 0.1\sin(s) - 0.1s$ and $g_2 = 0.1\sin(\Delta v) - 0.1\Delta v$ for Case~1. It is worth noting that the attacks shown in Case~1 -- Case~3 are taken from the set $\mathcal{C}$, while those of Case~4 -- Case~6 are drawn from the complement of $\mathcal{C}$.}}\label{speed_profile}
\end{figure}

Here we investigate the influence of attacks, taken from the set $\mathcal{C}$, on traffic flow dynamics. Furthermore, we explore the effects of attacks drawn from the complement of set $\mathcal{C}$, understanding that such attacks can induce markedly abnormal driving behavior and potentially result in rear-end collisions, discussed in Remark~\ref{remark3.1}. For illustrative purposes, all vehicles are initially assumed to travel at a constant speed, $v^* = 21$~m/s, before any attacks occur, with the simulation duration covering $I = \left[0, 200\right]$~sec. In the Baseline case, no attacks take place, and vehicles maintain a constant speed. As shown in Table~\ref{Table_parameters_IDM}, the maximum vehicle acceleration is set at 1.4~m/s\textsuperscript{2}, whereas the maximum deceleration value is chosen to be 2.5~m/s\textsuperscript{2}, slightly higher than the comfortable deceleration, 2.0~m/s\textsuperscript{2}. The free-flow speed is 30~m/s, matching the desired speed $v_0$ as indicated in Table~\ref{Table_parameters_IDM}. Following the literature, these parameter values are adopted only for the purposes of numerical illustration.

\begin{figure}[t!]
  \centering
  \subfloat[Case~1]{\includegraphics[width=0.33\textwidth]{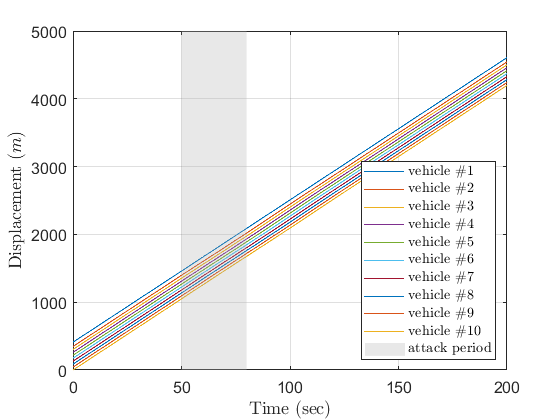}\label{displacement_case1}}
  \hfil%
  \subfloat[Case~2]{\includegraphics[width=0.33\textwidth]{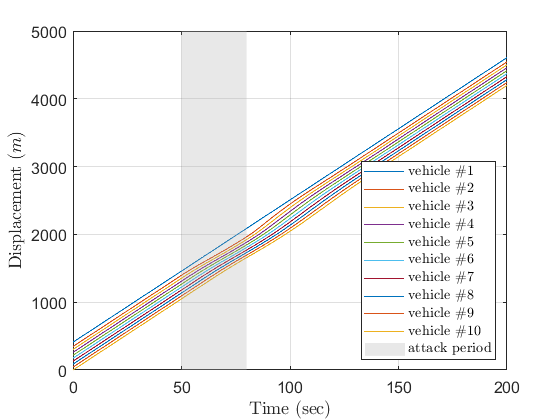}\label{displacement_case2}}
  \hfil%
  \subfloat[Case~3]{\includegraphics[width=0.33\textwidth]{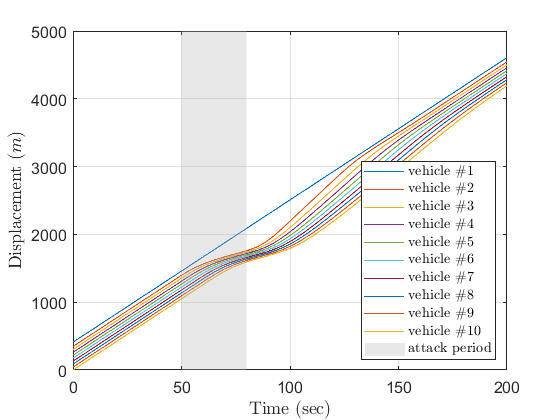}\label{displacement_case3}}
  \vfil%
  \subfloat[Case~4]{\includegraphics[width=0.33\textwidth]{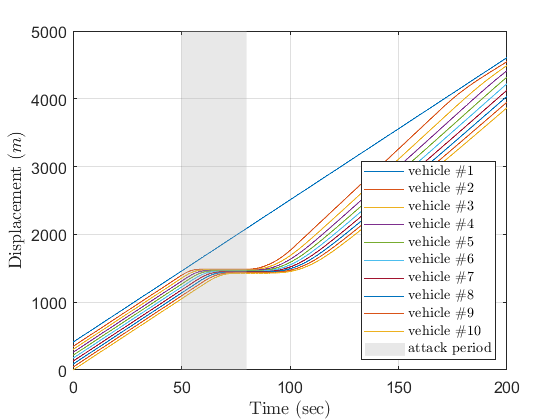}\label{displacement_case4}}
  \hfil%
  \subfloat[Case~5]{\includegraphics[width=0.33\textwidth]{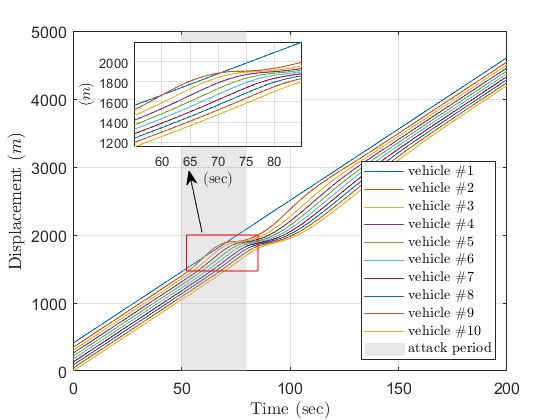}\label{displacement_case5}}
  \hfil%
  \subfloat[Case~6]{\includegraphics[width=0.33\textwidth]{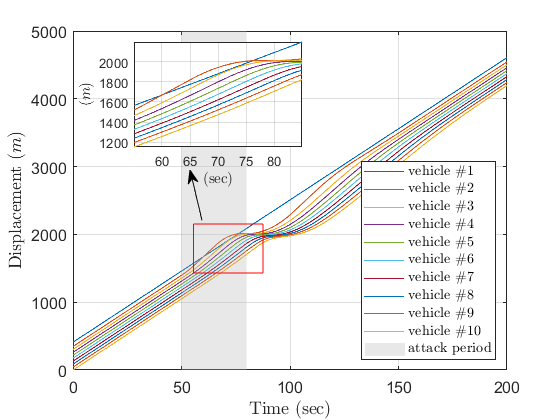}\label{displacement_case6}}
  \caption{\textnormal{Displacement profile of all vehicles for Case~1 -- Case~6, with attacks occurring during the period of $I = \left[50, 80\right]$~sec. The simulation settings of each case are same as those of that case shown in Figure~\ref{speed_profile}.}}\label{displacement_profile}
\end{figure}

As shown in Figure~\ref{Simulation_setting}, the ACC vehicle (\#2) is attacked over the period of $\left[50, 80\right]$~sec. We first examine three distinct cases with attacks taken from $\mathcal{C}$, specifically Case~1: $\omega_{1,i} = 0.1\sin(s_i) - 0.1s_i$ and $\omega_{2,i} = 0.1\sin(\Delta v_i) - 0.1\Delta v_i$, Case~2: $\omega_{1,i} = 0.4\sin(s_i) - 0.5s_i$ and $\omega_{2,i} = 0.4\sin(\Delta v_i) - 0.5\Delta v_i$, and Case~3: $\omega_{1,i} = - 0.9s_i$ and $\omega_{2,i} = - 0.9\Delta v_i$. Clearly, these attacks satisfy the conditions shown in equation~\eqref{eq3.19}. The resulting speed profiles are presented in Figure~\ref{speed_case1} through Figure~\ref{speed_case3}, where the attack period is highlighted in grey. An examination of these figures reveals that the ACC vehicle encounters speed disturbances as a consequence of malicious attacks. These disturbances also propagate to the vehicles behind the attacked ACC vehicle, persisting well beyond the attack period. Notably, disruptions to the bulk traffic become increasingly pronounced as the severity of the attacks escalates. The corresponding outcomes regarding vehicle displacement (trajectory) can be found in Figure~\ref{displacement_case1} through Figure~\ref{displacement_case3}, illustrating the emergence of shockwaves that impede traffic flow. This is more visible in Figure~\ref{displacement_case3} with more severe attacks, compared to Figure~\ref{displacement_case1} and Figure~\ref{displacement_case2} with much subtle attacks. 

So far, our analysis has focused on candidate attacks originating from $\mathcal{C}$, which could be challenging to detect and identify. We now examine Remark~\ref{remark3.1} considering attacks taken from the complement of $\mathcal{C}$ in Cases 4, 5, and 6. The corresponding vehicle speed and displacement are shown in Figure~\ref{speed_case4} through Figure~\ref{speed_case6}, and Figure~\ref{displacement_case4} through Figure~\ref{displacement_case6}, respectively. In Case~4, it is observed from Figure~\ref{speed_case4} that the attacked vehicle inexplicably exhibits a complete stop, causing a disruptive bottleneck for the vehicles behind, as observed in Figure~\ref{displacement_case4}. In both Case~5 and Case~6, the attacked ACC vehicle exhibits irrational driving behavior (Figure~\ref{speed_case5} and Figure~\ref{speed_case6}), resulting in deadly consequences of rear-end collisions, as seen in Figure~\ref{displacement_case5} and Figure~\ref{displacement_case6}, due to attacks originating from outside of $\mathcal{C}$. For attacks to remain stealthy, it is more reasonable to launch them from the set $\mathcal{C}$, as those originating from its complement can lead to conspicuously abnormal driving behavior and even collisions.

\subsection{Impact of candidate attacks on average speed variation}\label{section4_2}

Although attacks originating from $\mathcal{C}$ typically maintain a higher degree of stealthiness by inducing subtle changes in ACC driving behavior, they can still result in extensive disruption to the transportation system. Consequently, we investigate the potential impact of these malicious yet stealthy attacks on traffic performance using average speed variation (ASV) of vehicles. The ASV is defined per vehicle per second and can be used to evaluate traffic smoothness. Generally, a smaller ASV indicates smoother traffic flow, thereby higher efficiency. Specifically, the ASV over a given period of time, e.g., $\left[t_1, t_2\right]$, is given by~\citep{wang2023general}
\begin{eqnarray}
\textnormal{ASV} = \frac{1}{N(t_2 - t_1)}\sum_{i}\int_{t_1}^{t_2}\left|v_i(t) - v^{*}\right|dt,   \label{eq4.1}
\end{eqnarray}
where the index $i$ starts from the first ACC vehicle being attacked in the platoon since vehicles ahead of it do not experience any ripple effect due to attacks; clearly, the index $i$ goes from 2 to 10 as seen in Figure~\ref{Simulation_setting}. $N$ is the total number of upstream vehicles starting from the first attached ACC vehicle, i.e., $N = 9$. In other words, the ASV corresponding to Figure~\ref{Simulation_setting} is calculated for vehicles \#2 -- \#10 with $N = 9$ and the index $i$ starting from 2 in equation~\eqref{eq4.1}. Since attacks start to occur at $t = 50$~sec, it is reasonable to set $t_1 = 50$~sec, while $t_2 = 200$~sec same as the end of the simulation horizon. 

\begin{figure}[t!]
    \centering
    \includegraphics[width=0.5\textwidth]{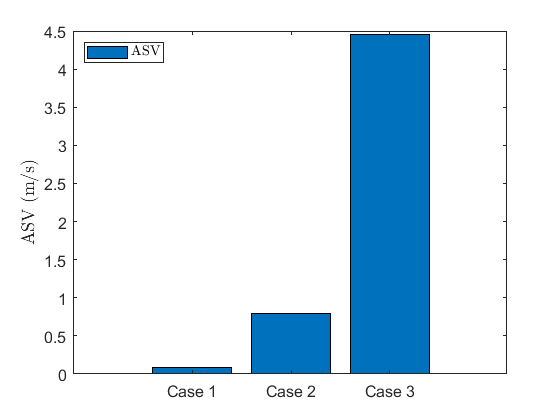}
    \caption{\textnormal{ASV of Case~1, Case~2, and Case~3, with attacks launched from the set $\mathcal{C}$. The value of ASV is calculated for vehicles \#2 -- \#10 over the period of $\left[50, 200\right]$~sec.}}
    \label{ASV_comparison}
\end{figure}

The results on ASV are presented in Figure~\ref{ASV_comparison} for Case~1, Case~2, and Case~3, with attacks drawn from $\mathcal{C}$. While some stealthy attacks may have minor impacts on traffic performance by altering ACC driving behavior in a highly subtle manner (Case~1), others could cause a fairly significant disruption to traffic flow as seen in Case~2 and Case~3 of Figure~\ref{ASV_comparison}. Clearly, the negative impact of malicious attacks on traffic efficiency in terms of ASV increases with the increase of attack severity, i.e., Case~1 $<$ Case~2 $<$ Case~3. 

\subsection{Impact of candidate attacks on energy efficiency}\label{section4_3}

As shown above, even subtle attacks drawn from the set $\mathcal{C}$ can cause considerable disruptions to traffic flow in terms of ASV. Here we assess the impact of such attacks on vehicular energy consumption. We employ the extensively used VT-Micro model~\citep{ahn2002estimating} for the estimation of fuel consumption. It is a regression model that utilizes instantaneous vehicle speed and acceleration as its input variables. The measure of effectiveness (MOE) is defined as~\citep{ahn2002estimating}
\begin{numcases}{\ln(\text{MOE}) =}
    \sum_{p=0}^{3}\sum_{q=0}^{3}(L_{p,q} \times v^{p} \times u^{q}), u \geq 0    \label{eq4.2a}  \\  
    \sum_{p=0}^{3}\sum_{q=0}^{3}(M_{p,q} \times v^{p} \times u^{q}), u < 0    \label{eq4.2b}
\end{numcases}
The MOE is the instantaneous fuel consumption rate~(L/s); $v$ and $u$ denote the instantaneous vehicle speed~(km/h) and acceleration~(km/h/s), respectively; $L_{p,q}$ and $M_{p,q}$ signify the regression coefficients for positive and negative acceleration, respectively. Interested readers are referred to~\citep{ahn2002estimating} for an in-depth discussion of the model, with further details about equations~\eqref{eq4.2a}--\eqref{eq4.2b} and its corresponding parameter values.

\begin{figure}[t!]
    \centering
    \includegraphics[width=0.5\textwidth]{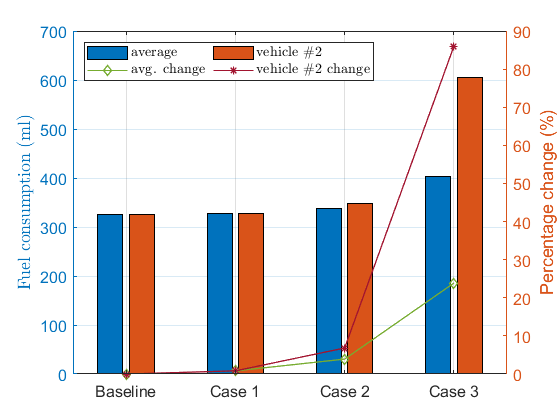}
    \vspace*{-2mm}
    \caption{\textnormal{Comparison of fuel consumption: Baseline vs. Case~1, Case~2 and Case~3. Blue bars show the average fuel consumption of vehicles \#2 -- \#10, while red bars denote that of vehicle \#2, over the period of $I$. The green and red lines represent the percentage increase in average fuel consumption of vehicles \#2 -- \#10 and that of vehicle \#2, respectively.}}
    \label{fuel_consumption}
\end{figure}

Figure~\ref{fuel_consumption} illustrates the fuel consumption of the compromised vehicle \#2 alone (red bars) and the average fuel consumption of vehicles \#2 -- \#10 (blue bars) over the period of $I$. As mentioned previously, in the Baseline case, all vehicles maintain a constant speed of 21~m/s, and no attacks are initiated, resulting in all vehicles exhibiting uniform fuel consumption levels. For Cases~1--3, both the fuel consumption of vehicle \#2 and the average fuel consumption of vehicles \#2 -- \#10 surpass their Baseline counterparts due to the adverse effects of attacks. Figure~\ref{fuel_consumption} also shows that as the attack severity increases (Case~1 $<$ Case~2 $<$ Case~3), the fuel consumption of the attacked vehicle and the average fuel consumption of all vehicles rise, with the percentage changes from the Baseline represented by the red and green lines, respectively. Clearly, even when designed to remain stealthy, attacks originating from $\mathcal{C}$ can still yield substantial negative energy consequences for both the targeted vehicle and the overall bulk traffic, as exemplified in Case~2 and Case~3.

\section{Conclusions and Discussion}\label{section5}

As the number of ACC vehicles continues to grow in the auto market, it also opens up opportunities for potential malicious attacks. In this study, we have presented an analytical framework designed to model and synthesize false data injection attacks on ACC vehicles, without relying on specific probability distributions. This framework promises to offer valuable insights for devising effective strategies aimed at detecting and mitigating these attacks. Leveraging this mathematical framework, we characterize the malicious and covert attributes of candidate attacks, providing a physical interpretation rooted in the understanding of car-following behavior. For technical completeness, we have considered attacks that could be launched in an additive as well as in a multiplicative fashion. To illustrate the modeling mechanism, we have conducted a series of numerical simulations to holistically assess the influence of attacks on car-following dynamics, average vehicle speed variation, and vehicular fuel consumption. The primary findings are very interesting in the sense that one can strategically synthesize candidate attacks that alter the driving behavior of ACC vehicles in a subtle manner yet result in significant disruptions to the traffic flow. Importantly, we have provided a series of analytical results to support the numerical illustration. 

Following many prior studies modeling AVs, we have adopted the widely used OVRV model to characterize the car-following dynamics of ACC vehicles. While empirical evidence indicates that the OVRV model is a good fit for both simulated and actual vehicle trajectories featuring ACC functionality, it is still not exactly clear how highly automated vehicles are to be operated in the future. As a result, one may need to introduce certain adaptation to the existing models for intelligent vehicles with a varying degree of automation as increasingly advanced AV technologies emerge. While the mathematical framework developed covers a broad range of candidate attacks, other forms of potential attacks are also worth looking into, especially when taking into account V2V communication in the context of CACC.

This study is the first of its kind in analytically characterizing malicious yet stealthy candidate attacks targeting ACC vehicles with explicit consideration of vehicle driving behavior. While it is hardly possible to exhaustively obtain the exact mathematical form of all potential attacks, the present work provides an effective method for synthesizing candidate attacks and offers useful insights into understanding their realistic impacts on future transportation systems, which motivates the development of efficient and robust attack detection and mitigation strategies in future studies, including car-following anomaly detection and advanced vehicle control design. Moreover, from the attacker perspective, one may attempt to maximize the disruption to traffic flow, which we believe can be appropriately formulated as an optimization problem under the developed framework with the conditions on attack serving as part of the constraints. An interesting question that naturally arises is how to alleviate the resulting impact of such intelligent attacks, which is of significant interest for a follow-up study.

\bibliographystyle{unsrtnat}
\bibliography{TRR_bibliography}

\end{document}